\documentclass[12pt,preprint]{emulateapj}
\usepackage{lscape}
\shorttitle{[\ion{O}{3}]$\lambda$ 4363 in the distant universe.}
\shortauthors{Hoyos \textit{et al.}}
\begin{document}

\title{The DEEP2 Galaxy Redshift Survey: Discovery of  Luminous, Metal-poor, Star-forming Galaxies at Redshifts $z \sim 0.7$\altaffilmark{1}}

\author{C. Hoyos \altaffilmark{2}\email{carlos.hoyos@uam.es} 
D. C. Koo \altaffilmark{3}\email{koo@ucolick.org} 
A. C. Phillips  \altaffilmark{3}\email{phillips@ucolick.org}
C. N. A. Willmer \altaffilmark{3}\email{cnaw@ucolick.org}
P. Guhathakurta \altaffilmark{3}\email{raja@ucolick.org}}

\altaffiltext{1}{Based on observations obtained with
the KECK2 Telescope at the W. M. Keck Observatory, which is operated as a scientific 
partnership among the California Institute of Technology, the University of 
California, and the National Aeronautics and Space Administration. The Observatory 
was made possible by the generous financial support of the W.. M. Keck Foundation.
This work also uses data obtained with the NASA/ESA {\em Hubble Space Telescope} through the Space Telescope
Science Institute,  which is operated by the Association of Universities for
Research in Astronomy (AURA), Inc., under NASA contract NAS 5-26555.}
\altaffiltext{2}{Departamento de F\'{\i}sica Te\'orica (C-XI), Universidad
Aut\'onoma de Madrid. Carretera de Colmenar Viejo km 15.600 28049 Madrid, Spain.}
\altaffiltext{3}{UCO/Lick Observatory and Department of Astronomy and Astrophysics, University of California Santa
Cruz, Santa Cruz, CA 95064.}

\textbf{To appear in Astrophysical Journal Letters.}

\begin{abstract}

We have discovered a sample of 17 metal-poor, yet luminous,
star-forming galaxies at redshifts $z \sim 0.7$.  They were selected
from the initial phase of the DEEP2 survey of 3900 galaxies and the Team Keck
Redshift Survey (TKRS) of 1536 galaxies as those showing the
temperature-sensitive [\ion{O}{3}]$\lambda4363$ auroral line.  These
rare galaxies have blue luminosities close to $L^{*}$, high star
formation rates of 5 to 12 $M_{\odot}$yr$^{-1}$, and oxygen abundances of
1/3 to 1/10 solar. They thus lie significantly off the
luminosity-metallicity relation  found previously for field
galaxies with strong emission lines at redshifts $z \sim 0.7$.  The
prior surveys relied on indirect, empirical calibrations of the
$R_{23}$ diagnostic
and the {\it assumption\/} that luminous galaxies are {\it not\/}
metal-poor.  Our discovery suggests that this assumption is sometimes
invalid.
As a class, these newly-discovered galaxies are: (1)~more metal-poor than
common classes of bright emission-line galaxies at $z \sim 0.7$ or at the
present epoch; (2)~comparable in metallicity to $z \sim 3$ Lyman Break
Galaxies but less luminous; and (3)~comparable in metallicity to local
metal-poor eXtreme Blue Compact Galaxies (XBCGs), but more luminous.
Together, the three samples suggest that  
the most-luminous, metal-poor, compact
galaxies become fainter over time.

\end{abstract}


\section{Introduction.}

The metal content of galaxies is an important diagnostic because it
relates directly to the integral history of star formation, galaxy
mass, and the inward and outward flows of gas 
\citep[see reviews by][ or, for a review on chemical evolution models, see \citealp{cen_ost_99}] 
{ostlin00, pagel97}.  Local 
studies reveal the existence of a luminosity-metallicity relation (LZR)
\citep{lequeux79,skill89_1,kindav81,rich_call95,camposa93} that  
presumably arises from the higher retention rate of enriched gas
in the gravitational wells of galaxies with larger masses, where the
assumption is that more luminous galaxies are also more massive.  The 
luminosity-metallicity relation (LZR) is expected to evolve over the lifetime of galaxies, but any predicted changes
in the slope, offset, and dispersion of the LZR are subject to many uncertainties.
Observations of the metallicity of galaxies at intermediate
redshifts $z > 0.5$ have been few and include three studies of field
galaxies at $z \sim $ 0.5 to 1 by \citet[henceforth K03]{kob03}, \citet[henceforth L03]{cfrs_lilly}, 
and \citet[henceforth KK04]{koke04}  and a few targets at very
high redshifts $z \sim 2.5$ by, e.g.,  \citet{pett01} and  \citet[henceforth KK00]{kob_koo_2000}.
The intermediate-redshift studies suggest that, at a given metallicity, galaxies
were typically more luminous in the past, while
the high-redshift samples show metallicities that are  sub-solar with
luminosities 5--40 times brighter than local galaxies of comparable metallicity.

The distant galaxy metallicities in these studies were all based on the 
[O/H]\footnote{We will henceforth refer to 12+log(O/H) as ``[O/H]''.} of the
emission lines and estimated from the empirical $R_{23}$ method introduced
by \citet{pagel_r23}, and further developed by \citet{mcgaugh91} and
\citet{py00}, among others. No galaxies had less than 1/3 solar
abundances, but this was in part due to the \emph{assumption} of using the
metal-rich (upper) branch of the $R_{23}$-metallicity relation.  This
letter presents a new sample of distant galaxies selected for the presence
of the [\ion{O}{3}]$\lambda$ 4363 \AA \ auroral line. This line is sensitive to
electron-temperatures \citep{ost89} and can, together with
$H_{\beta}$ and other oxygen lines, provide reliable gas metallicities
without assumptions about the ionization and metallicity.
This  selection also strongly favors [O/H] abundances less than $\sim 1/3$
solar and has enabled us to discover a new distant sample of luminous metal-poor
field galaxies.  We summarize our observations and
measurements in \S2; we present our data analysis in \S3 and compare our results
to the LZR derived from previous studies of field galaxies.  The main conclusions of this
study are presented in \S4.

We adopt the concordance cosmology, i.e., a flat Universe
with $\Omega_{\Lambda} = 0.7$ and $h = 0.7$. Magnitudes are all on
the Vega system.

\section{Observations \& Measurements}
\label{samp_met}

Galaxies were selected by inspection of reduced spectra from two 
redshift surveys of faint field galaxies, DEEP2 and TKRS, both using
the DEIMOS spectrograph
\citep{faber03} on the 10-meter Keck II Telescope.  DEEP2
\citep{deep_deimos} spectra had a total exposure time of one hour, covered the
wavelength range $\sim$ 6400-9000 \AA \ with the 1200 mm$^{-1}$
grating, and yielded FWHM resolutions of around 60 km s$^{-1}$. The initial
DEEP2 sample consisted of 3900 galaxies, 1200 of which had redshifts that
allowed the [\ion{O}{3}]$\lambda 4363$ and 
[\ion{O}{3}]$\lambda 4959$ lines in principle to be observed. This search
yielded 14 galaxies, or about 1\%, that display the weak auroral line
[\ion{O}{3}]$\lambda 4363$ along with prominent oxygen emission lines.

\begin{figure}
\plotone{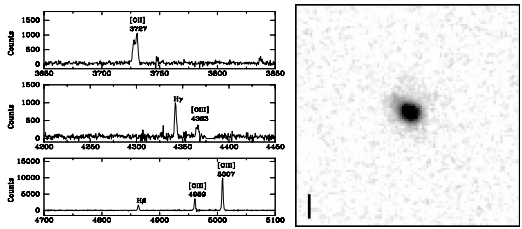}
\caption{
Spectrum of a low-metallicity (1/10 solar) galaxy at redshift $z = 0.68$ (TK-2 in Table 1) 
showing the temperature-sensitive [\ion{O}{3}]$\lambda$4363  line used to identify the sample and the 
other lines used to measure the gas phase abundance [O/H]. From top to bottom, (i) the [\ion{O}{2}]$\lambda$3727 line, (ii) the
 H$\gamma$ and [\ion{O}{3}]$\lambda$4363 lines,  and (iii) the  $H_{\beta}$ and 
[\ion{O}{3}]$\lambda, \lambda$4959,5007 lines. The $HST$ ACS image is taken in the $F814W$ filter 
(close to rest frame $B$);  North is up, and East is to the left. The image is 
3 \arcsec $\times$ 3\arcsec (18 kpc $\times$ 18 kpc). The half-light radius of this galaxy is 0.7 kpc. The thick, dark
line shown represents 2 kpc.
\label{spectrum_image}
}
\end{figure}

The TKRS \citep{wirth04}  is a
similar one hour survey targeting the GOODS-North field \citep{giava04}.
It used the 600 mm$^{-1}$ grating, and covered a wider range of wavelengths
(4600--9800 \AA), but had a lower FWHM resolution of 170 km s$^{-1}$. 
This survey yielded  1536 galaxies with reliable redshifts.
For 1090 galaxies, the redshifts 
allowed the [\ion{O}{3}]$\lambda 4363$ and 
[\ion{O}{3}]$\lambda 4959$ lines to be observable.
Of these, three galaxies, or 0.3\%, showed the auroral line and had 
redshifts above $z \sim 0.5$.
Fig. \ref{spectrum_image} shows one example along with its  HST
image. Table \ref{tabla} identifies all 17 targets, henceforth called
the O-4363 sample and tabulates the measurements 
described below.

The [O/H] metallicities are derived from emission lines,
including the temperature sensitive [\ion{O}{3}]$\lambda 4363$ line
along with [\ion{O}{2}]$\lambda3727$, $H_{\gamma}$, $H_{\beta}$  and
[\ion{O}{3}]$\lambda,\lambda 4959,5007$.  For the DEEP2 sample, only
4/14 galaxies possessed the full set of lines,  while 10 had
the [\ion{O}{2}]$\lambda 3727$ lines outside the observable
wavelength range. All oxygen lines were detected for the 3 TKRS
galaxies. When [\ion{O}{2}]$\lambda 3727$ was unobservable, its  line strength was 
estimated using the following fit to local \ion{H}{2} galaxies, with errors
about 50\% larger than from using direct [\ion{O}{2}] measurements (A. D\'{\i}az, private communication):
\begin{equation}
\log \frac{\mathrm{[OIII]}}{\mathrm{[OII]}}= (0.877\pm0.042)\times \log EW(H\beta) -1.155 \pm 0.078
\end{equation}
\noindent 
The electron temperature in the [\ion{O}{2}] zone was then derived
according to the method given in \cite{epm_diaz},  while the oxygen
abundances were all calculated using the formulae given in
\cite{pagel92.orig}. Objects showing [\ion{O}{2}]$\lambda$ 3727
have abundance uncertainties set to 0.1 dex, while the others
have uncertainties of 0.15 dex.

Blue absolute magnitudes ($M_{B}$) and rest-frame \ub \ colors were calculated 
from the \textit{BRI} photometry \citep{coil04} in DEEP2 and the 4-band \emph{HST-ACS} photometry
in the GOODS-N field of the TKRS, with K-corrections following 
those described by Willmer et al. (2005). Half-light
radii, $R_e$, were estimated from curve-of-growth profiles derived
from multi-aperture photometry of the \emph{HST ACS} image taken with the filter 
that yielded the closest match to restframe $B$ at the target's redshift. 
The star formation rates (SFR) were calculated from the $H_{\beta}$ luminosity
as in \citet{ken94}, valid for T$_{e}=10^{4}K$ and case B
recombination.
Since the DEIMOS spectra are not flux calibrated,
the $H_{\beta}$ line luminosity was estimated via  $M_{B}$ and $EW(H_{\beta})$  following  \cite{jmelnickxx}, 
with no extinction or color corrections. The derived luminosities and SFR are thus lower limits.

\begin{deluxetable*}{llllllllllll}
\tabletypesize{\footnotesize}
\tablecolumns{12}
\tablewidth{0pc}
\tablecaption{[\ion{O}{3}]$\lambda$4363 Selected Galaxies.\label{tabla}}

\tablehead{
\colhead{ID}\tablenotemark{a} & \colhead{z}\tablenotemark{b} & 
\colhead{RA} & \colhead{DEC} &
\colhead{12+log(O/H)} & \colhead{[\ion{O}{3}]/H$\beta$} &
\colhead{EW(H$\beta$)} & \colhead{M$_{B}$}\tablenotemark{c} & 
\colhead{\ub}\tablenotemark{d} & \colhead{SFR} &
\colhead{Vel. $\sigma$} & \colhead{$R_{e}$}\tablenotemark{e}\\
\colhead{\nodata} & \colhead{\nodata} & \colhead{(J2000)} & \colhead{(J2000)} &
\colhead{\nodata} & \colhead{\nodata} & \colhead{Rest frame \AA} & \colhead{\nodata} &
\colhead{(mag)} & \colhead{(M$_{\odot}$yr$^{-1}$)} & \colhead{(km-s$^{-1}$)} &
\colhead{(kpc)}
}

\startdata

D$^{2}$-1  & 0.851* & 23 29 08.20 & +00 20 40.70 & 8.1$\pm$0.1 & 8.0$\pm$0.8 & 98$\pm$5   &  -19.90 & -0.47   & 11.5$\pm$0.6 & 40$\pm$14 & \nodata \\
D$^{2}$-2  & 0.730* & 02 29 33.65 & +00 26 08.00 & 7.9$\pm$0.1 & 8.1$\pm$0.8 & 90$\pm$5   &  -19.24 & -0.47   & 5.8$\pm$0.3  & 25$\pm$6 & \nodata \\
D$^{2}$-3  & 0.749* & 16 53 03.49 & +34 58 48.90 & 7.8$\pm$0.1 & 6.3$\pm$0.6 & 96$\pm$5   &  -19.23 & -0.35   & 6.1$\pm$0.3  & 30$\pm$14 & \nodata \\
D$^{2}$-4  & 0.631  & 23 28 47.84 & \nodata & 8.0$\pm$0.15 & 6.1$\pm$0.6 & 88$\pm$5   &  -19.27 & -0.51   & 5.8$\pm$0.3  & 32$\pm$4 & \nodata \\
D$^{2}$-5  & 0.636  & 23 28 41.65 & +00 18 20.00 & 8.2$\pm$0.15 & 8.1$\pm$0.8 & 96$\pm$5   &  -19.24 & -0.34   & 6.2$\pm$0.3  & 28$\pm$4 & \nodata \\
D$^{2}$-6  & 0.530  & \nodata & \nodata & 8.1$\pm$0.15 & 7.8$\pm$0.8 & 160$\pm$20 &  \nodata & \nodata  & \nodata & 34$\pm$5 & \nodata \\
D$^{2}$-7  & 0.706  & 16 50 05.43 & +35 06 30.40 & 8.1$\pm$0.15 & 7.4$\pm$0.7 & 70$\pm$5   &  -20.10 & -0.35   & 9.8$\pm$0.7  & 41$\pm$4 & \nodata \\
D$^{2}$-8  & 0.659  & 02 31 17.32 & +00 37 28.20 & 8.1$\pm$0.15 & 6.6$\pm$0.7 & 70$\pm$10  &  -19.97 & -0.49   & 9.0$\pm$1.0  & 50$\pm$10 & \nodata \\
D$^{2}$-9  & 0.750* & 02 30 20.03 & +00 42 49.70 & 8.1$\pm$0.1 & 5.9$\pm$0.6 & 91$\pm$5   &  -20.55 & -0.49   & 20.0$\pm$1.0 & 40$\pm$10 & \nodata \\
D$^{2}$-10 & 0.657  & 02 28 38.46 & +00 28 52.30 & 8.3$\pm$0.15 & 4.9$\pm$0.5 & 60$\pm$10  &  -21.40 & -0.52   & 27.0$\pm$5.0 & 60$\pm$16 & \nodata \\
D$^{2}$-11 & 0.551  & 02 28 40.39 & +00 36 07.70  & 8.0$\pm$0.15 & 8.1$\pm$0.8 & 81$\pm$9   &  -18.48 & -0.45   & 2.6$\pm$0.3  & 20$\pm$20 & \nodata \\
D$^{2}$-12 & 0.680  & 02 28 45.05 & +00 41 32.80 & 8.2$\pm$0.15 & 9.4$\pm$0.9 & 110$\pm$15 &  -21.00 & -0.28  & 26.0$\pm$3.0 & 32$\pm$2 & \nodata \\
D$^{2}$-13 & 0.702  & 02 30 02.06  & +00 47 34.70 & 8.3$\pm$0.15 & 4.8$\pm$0.5 & 91$\pm$3   &  -19.94 & -0.55   & 11.2$\pm$0.4 & 30$\pm$16 & \nodata \\
D$^{2}$-14 & 0.725  & 02 30 12.32 & +00 36 52.50 & 7.8$\pm$0.15 & 7.1$\pm$0.7 & 59$\pm$5   &  -19.53 & -0.29   & 5.0$\pm$0.4  & 37$\pm$5& \nodata \\ \hline

TK-1       & 0.855* & 12 36 42.83 & +62 20 01.58 & 8.0$\pm$0.1 & 5.3$\pm$0.5 & 150$\pm$20 & \nodata  & \nodata  & \nodata & 57$\pm$6 & 1.0 \\   
TK-2       & 0.681* & 12 36 33.02 & +62 15 37.52 & 7.8$\pm$0.1 & 8.2$\pm$0.8 & 100$\pm$15  & -19.35   & -0.38    &  7.0$\pm$1.0 & 48$\pm$4 & 0.7 \\
TK-3       & 0.512* & 12 36 50.22 & +62 17 17.91 & 7.8$\pm$0.1 & 3.2$\pm$0.3 & 16$\pm$3   & -19.30    & -0.28   &  1.1$\pm$0.2 & 21$\pm$8 & 1.6 \\

\enddata

\tablenotetext{a}{The first 14 entries are DEEP2 sources. The last 3
entries are TKRS galaxies.}
\tablenotetext{b}{Redshifts marked with asterisks denote
galaxies for which all emission lines from [\ion{O}{2}]$\lambda$3727 to [\ion{O}{3}]$\lambda$5007 could be measured.}
\tablenotetext{c}{Rest-frame Johnson blue absolute magnitude.}
\tablenotetext{d}{Rest-frame Johnson \ub  \ color.}
\tablenotetext{e}{B-band half-light radii from \emph{HST-ACS} imaging.}

\tablecomments{Redshifts, J2000 coordinates, oxygen abundances, [\ion{O}{3}]/H$\beta$, H$\beta$ rest-frame equivalent widths
(EW$_{\beta}$), absolute magnitudes, colors, star formation rates, emission line width ($\sigma$) 
and B-band half-light radii of the studied sample.}
\end{deluxetable*}

\section{The Luminosity-Metallicity  Relation (LZR)}
\label{sci1}

\begin{figure}
\plotone{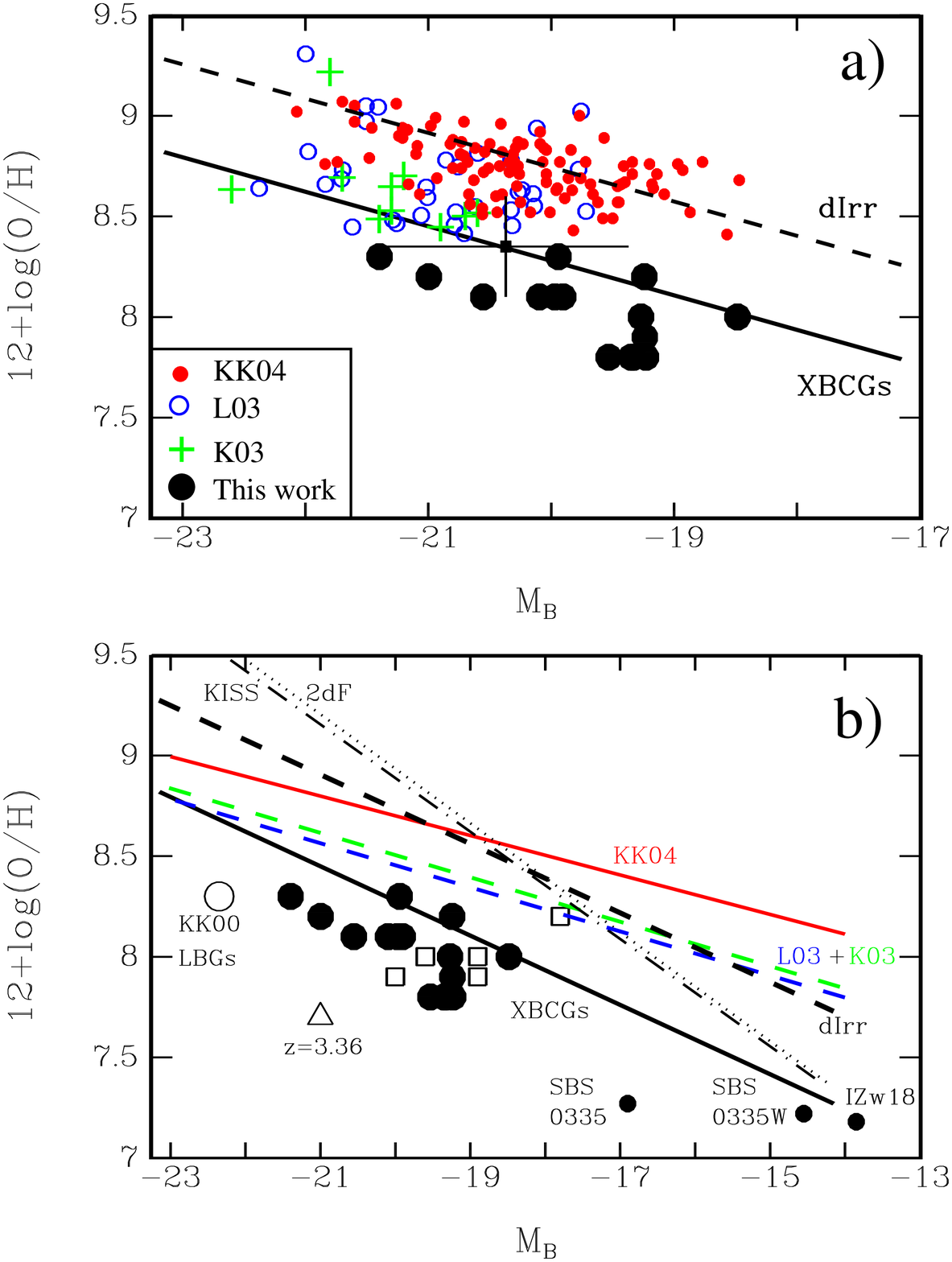}
\caption{
(a)~LZR diagram showing the intermediate-redshift samples as marked in the inset.
The [O/H] values for K03 and L03 were rederived using the 
\citet{py00} calibration, placing all surveys on the same system.  We estimate the errors to be  about 
0.2 dex for these data. In some fraction of cases, the resulting metallicities fell below 8.35, which is approximately
the limit between the high metallicity branch and the lower metalicity branch. However, given
the huge scatter of the $12+\log \mathrm{O/H}$-$R_{23}$ relationship in this regime ($\sim$0.4dex), they
might still be compatible with the use of the upper branch. For this reason, their oxygen abundance is fixed
at 8.35. The average position of these sources is given by the black solid square with error bars. The dashed dIrr line
is the average LZR  found for local dIrr   \citep{skill89_1, rich_call95} while the solid
XBCG line is the LZR for local,  metal-poor,  blue compact galaxies \citep{ostlin00}.~~~
(b)~LZ diagram showing possible local or  high-redshift  counterparts to the O-4363
sample. The three most metal-poor galaxies known are identified by
name.  Besides keeping the LZRs from panel~(a) of dIrr and XBCGs, we show
the LZR of two local,  emission-line galaxy samples: one from  KISS \citep{mel02} and the other from 2dF  \citep{lamareille04}. 
The big, open triangle is  for the $z = 3.36$ lensed galaxy  \citep{vi-mar04} and  the big open circle is
the average position of LBGs at $z \sim 3$ (KK00). Five local XBCGs from \citet{bergvost02} are shown
as open squares.  
\label{lz_diagram}
}
\end{figure}

The key result is seen in the [O/H] \textsl{vs.} $M_{B}$ relation in
Fig. \ref{lz_diagram}, which shows that [O/H]   
for the  17 galaxies in our O-4363 sample is 1/3 to 1/10 the solar 
value of [O/H]$_{\odot}$ = 8.69 \citep{allende01}.
While the O-4363 galaxies have  luminosities close to 
$L^{*}$ ($M_B \sim -20.4$ locally), they are offset to lower metallicities by about 0.6 dex 
in [O/H] when compared to 180 other $z \sim 0.7$ field galaxies studied by 
K03, L03, and KK04. All these studies used empirical calibrations, such as $R_{23}$, and 
adopted the upper, metal-rich branch\footnote{The K03 sample had 25 galaxies in the redshift 
range $0.60 < z < 0.81$;  L03 had  55 galaxies   between 0.48 and 0.91; and KK04 had 102 
between 0.55 and 0.85. }.

For galaxies  at $z\sim0.7$, the oxygen abundances derived here are the first using the direct method 
based on the temperature-sensitive  [\ion{O}{3}]$\lambda 4363$ line.  Our discovery
of  luminous galaxies with low [O/H] gas metallicities suggests that adopting 
the  metal-rich branch when using, e.g., the  $R_{23}$ 
method should be made with caution.  Such an assumption precludes finding  [O/H]  below $\sim$8.4.
If the empirical $R_{23}$ method and upper branch assumption were to be 
applied to the O-4363 sample,   [O/H] would be greater by about 0.4 $\pm$ 0.2 dex, nearly enough 
to place the O-4363 points atop the mean LZR of the $z \sim 0.7$ field galaxies (see Fig. 2).

\lastpagefootnotes

What fraction of the three other moderate-redshift samples are
actually metal-poor?  One estimate adopts two criteria suggested by
the O-4363 sample to identify metal-poor galaxies.  The first is based
on calculating $R_{23,0}$\footnote{Defined as the $R_{23}$ value that
an ionized \ion{H}{2} region would show if the reddening-corrected
ionization ratio [\ion{O}{3}]$\lambda, \lambda
4959,5007$/[\ion{O}{2}]$\lambda 3727$ were equal to one, leaving the
oxygen content unchanged.  We used the \citet{py00} calibration for
the upper branch.  In the case of the L03 objects, a uniform
extinction of $c(H\beta)=0.50$ was used. This value was adopted since this is the
average extinction found for emission-line galaxies in the Nearby Field Galaxy
Survey of similar luminosity \citep{cfrs_lilly}.
For both K03 and KK04 sources, EW's were used as surrogates for line strengths, but with no extinction
corrections for K03 and a uniform extinction of $c(H\beta)=0.40$ for
KK04. In this latter case, we have used the mean value of a very large sample
of bright local \ion{H}{2} galaxies from \citet{hoyos05}.} for all galaxies.  The O-4363 galaxies
have $R_{23,0} > 5$.  This places them near the turnaround region of the
$R_{23}$--[O/H] relation, where a small range in $R_{23}$ spans a wide
range in metallicity.  Our discovery of distant, luminous, metal-poor
galaxies in this region implies that the other distant samples may
also have such metal-poor galaxies.
The second criterion is based on large EWs of $H_{\beta}$. The O-4363 sample 
yield EW's greater than 40 \AA \ for all but one object
\footnote{TK-3 with TKRS catalogue ID-3653 has an unusually low
EW($H_{\beta}$) of about 20 \AA.  Its ionization ratio
[\ion{O}{3}]$\lambda,\lambda 4959+5007$/[\ion{O}{2}]$\lambda 3727$ is
approximately 0.7, and the ratio EW([\ion{O}{2}]$\lambda
3727$)/EW($H_{\beta}$) is $5\pm1$.  These values indicate that this
object is probably a Seyfert 2 galaxy, according to
\citet{rola97}. For all other objects for which the \citet{rola97}
diagnostics could be calculated, all tests indicate that they are
normal star-forming galaxies.}.  This additional criterion selects
those galaxies in the turnaround region that were most likely to be
metal-poor.  In the other distant galaxy surveys, we found 13 galaxies
(7\%) with high EW's of $H_{\beta}$ and $R_{23,0}$, that together
suggest low-metallicities.  This 7\% fraction is a lower limit since
some metal-poor galaxies may be outside the turnaround region or may
have smaller EW of $H_{\beta}$. In any case, independent tests are
critical to assess the true fraction of intermediate redshift galaxies
that have abundances below the upper branch, e.g., by observing
[\ion{N}{2}]/$H_{\alpha}$ in the near-infrared at our redshifts as
suggested by \citet{kewdop02} and \citet{denic02}.

What is the nature of our O-4363 sample, and do such galaxies exist
locally or at higher redshifts?  Fig. 2b. shows several relevant LZRs
from local to distant samples. One sees that relatively common samples
of emission line galaxies, such as those of local dIrr, the $z \sim 0.7$ 
galaxies from K03, L03, and KK04 and the local emission line
galaxies from 2dF or KISS surveys all have LZRs that are offset to metallicities
higher than that of the O-4363 sample\footnote{The comparison with the latter
samples of local galaxies should not be taken beyond $\mathrm{[O/H]}=9.0$, because
the KISS and 2DF abundances at high luminosities are clearly 
too high \citep{pett_pagel04}. It is then only below $\mathrm{[O/H]}=9.0$ 
($M_{B}\geq -20.5$) that valid comparisons can be made between our O-4363 sample
and the KISS or 2DF samples. Fortunately, most of the $z\sim 0.7$ O-4363 objects are less
luminous than this limit.}.
On the other hand, the O-4363 galaxies are far better matches to local XBCG's and even to the luminous Lyman Break
Galaxies (LBG) at redshifts $z \sim 2.5$ (KK00), or  to a
gravitationally-lensed galaxy at redshift $z = 3.36$, which has a
metallicity of 1/10 solar, a blue absolute magnitude of -21.0
 and a SFR of 6 $M_{\odot}$yr$^{-1}$ \citep{vi-mar04}\footnote{This object is rather
extreme, being 1.0 dex below L03, K03 and KK04 objects of similar luminosity.}. We do not find any 
correlation between the residuals of
the O-4363 sample with respect to the LZR of local Extreme Blue Compact
Galaxies (XBCG) \citep{ostlin00} with  \ub \ color, strength of
H$_{\beta}$, and internal 
velocity dispersion (see Table \ref{tabla}).
Much like the XBCG and LBG, the O-4363 galaxies
may belong to the compact class of galaxies. But this suggestion is
based presently on the small 1--2~kpc sizes seen in the only three
galaxies with \emph{HST} images. Moreover, the emission-line velocity
widths (see Table 1.) are narrow and suggest that the O-4363 galaxies
are more likely to be galaxies with small dynamical masses; the very blue colors and
high star formation rates  suggest a recent, strong burst of star 
formation. Overall,
the trend suggested by Fig. 2b is that the most luminous, metal-poor galaxies are getting fainter
with time.


\section{Summary.}\label{discu}

Based on a search for the [\ion{O}{3}]$\lambda 4363$ emission line in
the TKRS and initial DEEP2 surveys of field galaxies, we have
discovered 17 galaxies at redshift $z \sim 0.7$ that are luminous,
very blue, compact, and metal poor, roughly 1/3 to 1/10 solar in
[O/H]. Though rare, such metal-poor galaxies highlight the diversity
among galaxies with similar luminosities and serve as important
laboratories to study galaxy evolution \citep{ostlin00}. This sample
is lower in [O/H] by 0.6 dex on average in the LZR when compared to
prior studies at these redshift, which used empirical calibrations,
such as $R_{23}$.
The previous studies, however, assumed the metal-rich
branch of the calibration, while our results show that this assumption
may not apply, even for luminous galaxies, especially when high values
of EW(H$_{\beta}$) and $R_{23}$ are found (roughly 7\% of the other
samples).  Based on comparisons to local and high redshift samples, we
speculate that our metal-poor, luminous galaxies at $z \sim 0.7$
provide an important bridge between local  Extreme Blue
Compact Galaxies (XBCGs) and Lyman Break Galaxies (LBGs) at redshifts
$z \sim 3$. All three  samples share the property of being overluminous for their
metallicities, when compared to local galaxy samples, and of having very high EWs of $H_{\beta}$, typically above
40 \AA \ and up to 150 \AA \ and, thus, similar to that found for some
local \ion{H}{2} galaxies \citep{T91,hoyos05}.  The calculated star
formation rates of the O-4363 galaxies are mostly from 5 to 12
$M_{\odot}$yr$^{-1}$, indicating that the star-forming activity is
very strong (c.f., the SFR of 30 Doradus is 0.1 $M_{\odot}$yr$^{-1}$)
and thus lying roughly between that of local metal-poor, blue compact
galaxies and distant Lyman Break Galaxies. When DEEP2 is complete, we
expect to have a sample nearly 10$\times$ larger, many with \emph{HST}
images.  The resulting data set should thus provide vastly improved probes of
their nature, enable us to understand
their relationship to other classes of galaxies at different
epochs, and yield constraints on the physical processes involved in chemical and
galaxy evolution.

\acknowledgments

We thank A. I. D\'\i az and  R.. Guzm\'an  for useful discussions.
We acknowledge support from 
the Spanish DGICYT grant AYA-2000-0973, the MECD FPU grant AP2000-1389,
NSF grants AST 95-29028 and AST 00-71198,  NASA grant AR-07532.01-96,
and the New Del Amo grant. We close with thanks to the Hawaiian people for
allowing us to use their sacred mountain.


\end{document}